\documentclass[12pt,preprint]{aastex}
\newcommand{\be}{\begin{equation}}
\newcommand{\ee}{\end{equation}}

\newcommand{\vn}{{\bf v}_{n}}
\newcommand{\vns}{{\bf v}_{ns}}
\newcommand{\vs}{{\bf v}_{s}}

\newcommand{\vv}{{\bf v}}

\newcommand{\efe}{{\bf F}}

\newcommand{\pa}{\partial}
\newcommand{\rna}{\rho_{n}}
\newcommand{\rsa}{\rho_{s}}

\newcommand{\Rey}{{\it Re}}

\newcommand{\dEkman}{d_{\rm E}}
\newcommand{\vdg}{v_{\rm DG}}
\newcommand{\cmsec}{{\rm cm} \, {\rm s}^{-1}}
\newcommand{\radsec}{{\rm rad} \, {\rm s}^{-1}}
\def\ltsima{$\; \buildrel < \over \sim \;$}

\def\lsim{\lower.5ex\hbox{\ltsima}}
\def\gtsima{$\; \buildrel > \over \sim \;$}
\def\gsim{\lower.5ex\hbox{\gtsima}}

\begin{document}
\title{Transitions between turbulent and laminar 
superfluid vorticity states in the outer core of a neutron
star}

\author{C. Peralta,\altaffilmark{1,}\altaffilmark{2} A. Melatos,\altaffilmark{1} M. Giacobello\altaffilmark{3} and A. Ooi\altaffilmark{3} }

\email{cperalta@physics.unimelb.edu.au}

\altaffiltext{1}{School of Physics, University of Melbourne,
Parkville, VIC 3010, Australia}

\altaffiltext{2}{Departamento de F\'{\i}sica, Escuela de Ciencias,
Universidad de Oriente, Cuman\'a, Venezuela}

\altaffiltext{3}{Department of Mechanical and Manufacturing
Engineering, University
of Melbourne, Parkville, VIC 3010, Australia}


\begin{abstract}
\noindent We investigate the global transition
from a turbulent state of superfluid vorticity (quasi-isotropic vortex tangle)
to a laminar state (rectilinear vortex array), and vice versa, in
the outer core of a neutron star. By solving numerically
the hydrodynamic Hall-Vinen-Bekarevich-Khalatnikov
equations for a rotating superfluid 
in a differentially rotating spherical shell, we find that the meridional
counterflow driven by Ekman pumping
exceeds the Donnelly-Glaberson threshold throughout most
of the outer core, exciting unstable Kelvin waves which disrupt
the rectilinear vortex array, creating a vortex tangle. In the
turbulent state, the torque
exerted on the crust oscillates, and the crust-core coupling
is weaker than in the laminar state.
This leads to a new scenario
for the rotational glitches observed in radio pulsars: a vortex
tangle is sustained in the differentially rotating outer core by 
the meridional counterflow, a sudden spin-up event (triggered
by an unknown process) brings the crust
and core into corotation, the vortex tangle relaxes
back to a rectilinear vortex array (in $\lsim 10^5$ {\rm s}),
then the crust spins down electromagnetically until
enough meridional counterflow
builds up  (after $\lsim 1$ {\rm yr}) to reform a vortex tangle.
The turbulent-laminar transition can occur uniformly
or in patches; the associated time-scales are estimated from
vortex filament theory.
We calculate numerically the global structure of the flow with and without
an inviscid superfluid component, for Hall-Vinen (laminar) and Gorter-Mellink
(turbulent) forms of the mutual friction. We also calculate the post-glitch
evolution of the angular velocity of the crust and its
time derivative, and compare the results with radio pulse timing data,
predicting a correlation between glitch activity and Reynolds number.
Terrestrial laboratory experiments are proposed to test some of these ideas.
\end{abstract}


\section{Introduction}
Timing irregularities in a rotation-powered pulsar,
such as discontinuous glitches \citep{lss00,zwwmwz04,s05} and stochastic
timing noise \citep{hthesis02,sfw03}, provide an indirect probe of the
internal structure of the star. The physical processes usually invoked to 
explain these phenomena
are (un)pinning  of Feynman-Onsager vortices in the 
crystalline inner crust \citep{ai75}, starquakes \citep{r76}, and 
thermally driven vortex creep \citep{apas84,leb93}.
Less attention has been directed at the {\it global hydrodynamics}
 of the superfluid, except within the context of
the spin-up problem in cylindrical geometry
\citep{aprs78, r93, cls00}. The importance of
the global hydrodynamics was demonstrated by 
\citet{tsatsa80}, who simulated pulsar rotational irregularities 
in the laboratory by impulsively accelerating
rotating containers of He II, obtaining
qualitative agreement with radio timing data (e.g. glitch amplitudes
and post-glitch relaxation times).

In this paper, we examine how the global flow pattern of superfluid
in the outer core of a neutron star affects the rotation of the star. We
focus on the outer core for simplicity: vortex pinning is thought to be
weak or non-existent there \citep{als84,dp03,dp04}, and the fluid is mainly isotropic
($^1S_0$ Cooper pairing) \citep{ss95,yls99}, reducing the problem to a hydrodynamic
one in a spherical shell. Even with this simplification, the
calculation remains numerically challenging: the spherical Couette
problem for a superfluid was solved for the first time only recently
\citep{pmgo05a}, generalizing previous work on the cylindrical
Taylor-Couette problem for a superfluid \citep{hbj95,hb04} and the spherical
Couette problem for a classical viscous fluid \citep{mt87a,dl94}.

An isotropic ($^1S_0$-paired) neutron superfluid  is described by
the two-fluid Hall-Vinen-Bekarevich-Khalatnikov (HVBK) model.
In this model, the viscous normal fluid and inviscid superfluid 
components feel
a mutual friction force whose magnitude and direction depends on the 
distribution of Feynman-Onsager vortices \citep{hv56a,bk61}. 
The vortices are organized in a rectilinear array if the flow
is strictly toroidal, but they evolve into a tangle of
reconnecting loops when the counterflow along the rotation
axis exceeds a threshold, exciting the
Donnelly-Glaberson instability (DGI) \citep{gjo74,sbd83,tab03}.
\citet{pmgo05a} showed that the DGI is excited in a neutron
star under a wide range of conditions, driven by the meridional
component of the spherical Couette flow (SCF) in the interior.
The mutual friction force changes dramatically during
transitions between a vortex array and a vortex tangle
\citep{gm49,vinen57c,sbd83,s85}, affecting the rotational
evolution of the star.

In this paper, we propose a phenomenological model for timing irregularities
in radio pulsars based on the creation and destruction
of a vortex tangle --- {\it superfluid turbulence} \citep{bsd95,vinen03} --- in the 
outer core of a rotating 
neutron star. In our scenario, a glitch comprises
the following sequence of events. 
(i) Differential rotation between the outer core and
crust of the star, built up over time through electromagnetic spin down, 
generates a meridional Ekman counterflow in the outer core.
We show that the axial counterflow exceeds the DGI threshold,
creating a vortex tangle throughout the outer core. 
The mutual friction in this turbulent state 
takes the isotropic Gorter-Mellink (GM) form
and is much weaker than the mutual friction associated with a rectilinear
vortex array. (ii) When the glitch occurs,
triggered by an unknown mechanism, the outer core and inner crust
suddenly come into corotation and the vortex tangle decays, ultimately
converting into a rectilinear vortex array. The 
decay process lasts  $\lsim 10^5$ {\rm s}, depending on the drag
force acting on the vortex rings, after which the mutual
friction takes the anisotropic Hall-Vinen (HV) form \citep{hv56a,hv56b} and
increases by $\sim 5$ orders of magnitude,
precipitating a ``torque crisis". (iii) After
the glitch, differential
rotation builds up again between the outer core and the crust due to electromagnetic
spin down. When the axial counterflow exceeds the DGI threshold,
after $\lsim 1$ {\rm yr}, the vortex array breaks up again into a tangle
and the mutual friction drops sharply.
Similar transitions from
turbulent to laminar flow in a superfluid have been
observed in laboratory  experiments where He II, cooled to a few mK, flows
around an oscillating microsphere \citep{nks02,s04}.

The paper is organized as follows. HVBK theory
is briefly reviewed in \S\ref{sec:HVBKT}, together with the
pseudospectral numerical method which we use to solve the
HVBK equations in a differentially rotating spherical shell.
The physics of the turbulent-laminar
transition in a generic glitch scenario is elaborated in
\S\ref{sec:VTMODEL}.
The response of the stellar crust to
a turbulent-laminar transition in the outer core is calculated numerically in
\S\ref{sec:rotevol}. Finally, the results are summarized and applied
to observational data in \S\ref{sec:summary}; 
a fuller observational analysis will be carried out in a future paper.

\section{HVBK model of the outer core}
\label{sec:HVBKT}
The neutron superfluid in the outer core exists
in an isotropic ($^1S_0$) phase for densities $\rho$ in 
the range $0.6  < \rho/\rho_*  < 1.0 $  (where $\rho_*$
is the nuclear saturation density) and an anisotropic ($^3P_2$) phase for
$1.0  < \rho/\rho_*  < 1.6 $ \citep{yls99};
the depth at which the $^1S_0$-$^3P_2$ transition occurs
is not known precisely \citep{e88}.
As the hydrodynamic equations for the $^3P_2$ phase \citep{mm05} are complicated
and hard to handle numerically, we construct our model around the $^1S_0$
phase, described by HVBK theory in this paper. We 
plan to extend the model to the $^3P_2$ phase
in the future. 

The protons in the outer core are probably in a type II superconducting state,
where the magnetic field is quantized into fluxoids
 \citep{sauls89}. Protons inside magnetic
fluxoids interact
with neutrons in Feynman-Onsager vortices
\citep{sauls89,ruderman91c,l03}. If vortex pinning occurs, 
and if the core magnetic field has 
comparable
poloidal and toroidal components \citep{td93}, 
the geometry of the neutron-fluxoid
interaction can be complicated \citep{rzc98}.
To avoid these difficulties, we incorporate the protons
(and other charged species) into the normal component of the superfluid,
a common approximation \citep{cj03,pca04}.
We neglect proton-neutron entrainment \citep{m91a,ss95}, which is
observed in terrestrial $^3$He-$^4$He mixtures \citep{ab76,ac01}
and is an important mechanism modifying the mutual friction
for rectilinear vortices in a neutron star \citep{m91a,asc06}.
We neglect hydrodynamic forces
arising from entropy gradients and angular momentum textures \citep{mm05}.
Finally, we neglect
vortex pinning for simplicity, except at the inner
and outer boundaries, even though it seems likely that pinning plays
an important role in glitch dynamics \citep{eb88,bel92,hs01,lc02}.
\footnote{Three-fluid models analyzing post-glitch relaxation
favor pinning at the core boundary \citep{ss95}, e.g. due to vortex
cluster-Meissner current interactions \citep{sc99}.}

\subsection{HVBK theory}
\label{sec:HVBKT1}
The HVBK model for a rotating superfluid is a generalization 
of the Landau-Tisza two-fluid model that includes the hydrodynamic
forces exerted by quantized vortices in He II \citep{hv56a,bk61} and
$^1S_0$-paired neutron matter \citep{tt86,pca04}. Fluid
particles in the theory are assumed to be threaded by many
coaligned vortices, a valid assumption over length-scales longer than the average
inter-vortex separation (and shorter than the average
radius of curvature). In the continuum limit, the
vorticity of the superfluid component satisfies $\mbox{\boldmath$\omega$}_s = \nabla \times \vs \ne 0$
macroscopically (cf. $\nabla \times \vs = 0$ microscopically). 
The normal component
comprises charged species (protons and electrons, 
locked together by the external magnetic field) plus
thermal excitations and behaves like a classical, viscous,
Navier-Stokes fluid (kinematic viscosity $\nu_n$).
The isothermal, incompressible ($\nabla \cdot \vn
= \nabla \cdot \vs =0$) HVBK equations of motion take
the form \citep{bj88,hb00}
\begin{equation}
\label{eq:supeq1} \frac{d_n \vn}{dt} = -\frac{\nabla p_n}{\rho}
+ \nu_{n} \nabla^{2} \vn + \frac{\rsa}{\rho} {\bf F} -
\frac{\nu_s \rho_s}{\rho} \nabla {|\mbox{\boldmath$\omega$}_s|},
\end{equation}

\begin{equation}
\label{eq:supeq2}  \frac{d_s \vs}{dt} = -\frac{\nabla p_s}{\rho}
+ \nu_{s} {\bf T} - \frac{\rna}{\rho} {\bf F} - \frac{\nu_s \rho_s}{\rho} \nabla {|\mbox{\boldmath$\omega$}_s|},
\end{equation} with $d_{n,s} /dt = \pa /{\pa t} + \vv_{n,s} \cdot \nabla$,
where $\vn$ ($\vs$) and $\rho_{n}$ ($\rho_s$) are the normal
fluid (superfluid) velocities and densities respectively.
Effective pressures $p_{s}$ and $p_{n}$ are defined by $\nabla p_s
= \nabla p - \frac{1}{2} \rho_n \nabla (\vns^2)$ and $\nabla p_n =
\nabla p + \frac{1}{2} \rho_s \nabla (\vns^2)$, with $\vns =
\vn-\vs$. Note that, in neutron star applications, the gravitational
potential can be subsumed into $p$ without loss of generality,
in the incompressible limit.

The vortex tension force per unit mass, which arises from local 
circulation around 
quantized vortices \citep{am66}, is defined as 
\begin{equation}
\label{eq:tension}
\nu_{s} {\bf T}
= \mbox{\boldmath$\omega$}_{s} \times (\nabla \times \hat{\mbox{\boldmath$\omega$}}_{s}),
\end{equation}
with $\hat{\mbox{\boldmath$\omega$}}_{s} = \mbox{\boldmath$\omega$}_{s}/|\mbox{\boldmath$\omega$}_{s}|$. Here,
$\nu_{s} = (\kappa/4\pi) \ln(b_0/a_0)$ is the
stiffness parameter,  $\kappa = h/2 m_n$ is the quantum of
circulation, $m_n$ is the mass of the neutron, $a_0$ is the radius
of the vortex core, and $b_0$ is the intervortex spacing. Note
that $\nu_s$, which has the dimensions of a kinematic viscosity,
controls the oscillation frequency of Kelvin waves excited on
vortex lines \citep{hbj95}.

The mutual friction force  per unit mass, ${\bf F}$, arises from the
interaction between the quantized vortex lines and the normal
fluid (via roton scattering in He II and electron scattering in a
neutron star) \citep{hv56a,hv56b}. The structure of this force depends
on the global configuration of the vortices. If the
vortices form a rectilinear array, the friction
takes the anisotropic HV form \citep{hv56a,hv56b}

\begin{equation}
 \label{eq:ffhv}
 {\bf F} = \frac{1}{2} B \hat{\mbox{\boldmath$\omega$}}_{s} \times (
\mbox{\boldmath$\omega$}_{s} \times \vns - \nu_{s} {\bf T}) + \frac{1}{2}B^{\prime} 
(\mbox{\boldmath$\omega$}_{s}
\times \vns - \nu_{s} {\bf T})
\end{equation} where $B$ and $B'$ are temperature-dependent mutual friction
coefficients \citep{bdv83}. If the vortices form a vortex tangle, the
friction takes the isotropic GM form
\citep{gm49}
\begin{equation} \label{eq:gmforce} {\bf F} = A^{\prime} \left( \frac{\rho_{n}
\rho_{s} v_{ns}^{2}}{ \kappa \rho^{2}}\right) \vns, \end{equation} 
where
$A^{\prime} = B^3 \rho_n^2 \pi^2 \chi_1^2/3\rho^2 \chi_2^2$ is a
dimensionless temperature-dependent coefficient, related to the
original GM constant (usually denoted by $A$ in the literature)
by $A^{\prime} = A \rho \kappa$. Here, $\chi_1$ and $\chi_2$ are
dimensionless constants of order unity \citep{vinen57c}. 
The HV and GM forces are in the ratio $\sim 10^{5}$ under the
physical conditions prevailing in the outer core of a neutron star \citep{pmgo05a}. 
This implies that the normal and superfluid
components of the star (and hence the crust and core, through viscous torques) 
are effectively
uncoupled most of the time (when enough differential rotation
has built up), but become
tightly coupled in the immediate aftermath of a glitch when
a turbulent-to-laminar transition occurs \citep{pmgo05a}. It is 
important to bear this in mind when reading \S\ref{sec:VTMODEL}.

\subsection{Spherical Couette flow}
\label{sec:SCF}
We model the global hydrodynamics of the outer core
by considering an HVBK superfluid enclosed in a differentially
rotating spherical shell, i.e. superfluid spherical Couette flow. 
The inner (radius $R_1$) and
outer (radius $R_2$) surfaces of the shell rotate at angular
velocities $\Omega_1$ and $\Omega_2$, respectively, about a
common axis.
All quantities  are expressed in dimensionless form using $R_2$
as the unit of length and $\Omega_1^{-1}$ as the unit of time. We
adopt spherical polar coordinates $(r,\theta,\phi)$, with the $z$
direction along the axis of rotation.

The boundary conditions satisfied by the normal and superfluid
components are subtle.
Neither component can penetrate the boundaries, implying
$(\vn)_{r} = (\vs)_r = 0$. The tangential 
normal fluid velocity satisfies the no-slip Dirichlet condition, viz.
$(\vn)_{\theta} = 0$, $(\vn)_{\phi}= R_{1,2} \Omega_{1,2} \sin \theta$.
The behavior of the superfluid at the boundaries is influenced
by the quantized vortices, which
can either slide past, or pin to, irregularities on the boundaries.
The former scenario implies $(\mbox{\boldmath$\omega$}_{s})_\theta  =  (\mbox{\boldmath$\omega$}_{s})_\phi=0$,
while the latter implies $(\mbox{\boldmath$\omega$}_s)_r \mbox{\boldmath$\omega$}_s \times \vv_{L}= 0$
and requires a knowledge of the vortex line velocity $\vv_L$ locally.
As the HVBK model provides no information about $\vv_L$, we
adopt the widely
used no-slip compromise
$(\vs)_\theta=0$, $(\vs)_\phi = R_{1,2} \Omega_{1,2} \sin\theta$, which in the
spherical case does
not restrict the orientation of the
vortex lines at the surface \citep{hbj95,hb95,hb04}.
Note that, initially, $\vn$ and $\vs$
must be divergence-free, with
$\nabla \times \vs \neq 0$; the Stokes solution with
$\vs = \vn$ satisfies this constraint \citep{landau_fluidos}.
However, the numerical results presented in \S\ref{sec:VTMODEL} and
\S\ref{sec:rotevol} are obtained after
several Ekman times (and hence rotation periods) have elapsed, by which
time the flow has
forgotten its initial conditions, as in most astrophysical applications.

Many existing glitch models are based on
cylindrical geometries \citep{aprs78,apas84,ae96,ll02}, except
for a few experiments with spheres \citep{tsatsa73,tsatsa80}.
However, the global flow pattern in spherical Couette flow 
differs from its cylindrical counterpart in two important ways. 
First, in a cylinder,
the principal flow is toroidal, whereas, in a sphere,
the principal flow is an axisymmetric combination of toroidal flow
and meridional circulation for all $\Rey$ \citep{ttuckerman}. 
\footnote{Spherical Couette flow is a combination of
quasi-parallel-plate Ekman pumping at the pole and quasi-cylindrical,
centrifugally driven Taylor vortices at the equator \citep{je00}.}
Second, it is misleading
to approximate the equatorial region of a sphere by
a cylinder, even for $R_1 \approx R_2$; the curvature, although slight,
causes significant differences
in the critical Taylor number at which vortices appear \citep{sj83,stuart86}.
The differences between cylindrical and spherical Couette flow are equally
prominent in classical fluids and superfluids \citep{pmgo05a}.

Spherical Couette flow of a viscous fluid is controlled by three 
parameters: the Reynolds number
$\Rey = \Omega_1 R_2^2/\nu_n$, 
the dimensionless
gap width  $\delta = (R_2 - R_1)/R_2$,
and the angular shear $\Delta \Omega = \Omega_2 - \Omega_1$. 
If a superfluid is also introduced, $\nu_s$
emerges as an additional parameter. 
In the laminar regime ($\Rey \lsim 10^{4}$), 
the global flow 
can be classified according
to the number of cells of meridional circulation in each hemisphere.
In a narrow gap ($\delta < 0.11$),
the meridional circulation is slow and can be
approximated by 
$(\vn)_{\theta}/(R_2 \Delta \Omega) \approx 10^{-2} \delta^2 (R_2/R_1) \Rey$
\citep{yb86}.
As $\Rey$ increases, a centrifugal instability
generates toroidal Taylor vortices near the equator until, at
the onset of turbulence ($\Rey \gsim 10^{4}$), a helical traveling
wave develops.
In a wide gap ($\delta > 0.4$), by contrast,
all secondary flows are nonaxisymmetric, and they oscillate 
when $\Rey$ exceeds a threshold \citep{yb86}.
In classical fluids, these flows have been thoroughly investigated 
numerically
and experimentally \citep{ybm75,ybm77,bmy78,yb86,mt87a,mt87b,je00}.

Recently, high-resolution numerical simulations of 
superfluid spherical Couette flow were performed successfully
for the first time \citep{pmgo05a}.
\footnote{Visualizing superfluid spherical Couette flow experimentally
is challenging. It has been attempted successfully
using glass microspheres in
turbulent He II \citep{bs93}. The technique of
particle image velocimetry is currently being evaluated \citep{zcv04}.}
Figures \ref{fig:fig1}a and \ref{fig:fig1}b display streamlines
of the normal fluid and superfluid components for
$\Rey=3 \times 10^{4}$,
$\delta=0.3$, $\Delta \Omega = 0.1$,
and HV mutual friction.
Both components exhibit a similar structure in each hemisphere: 
a ``square" Ekman cell near the pole, three to four secondary
meridional cells, a number of smaller cells, 
and two equatorial vortices
near the outer boundary which 
are created and destroyed intermittently. We emphasize
that the normal component is not assumed to be in
uniform (solid body) rotation; it is affected by
the superfluid  (through mutual friction)
and the boundary conditions (differential rotation).
Figures \ref{fig:fig1}c and \ref{fig:fig1}d display
streamlines for a run with identical parameters 
and GM mutual friction, for which the inter-fluid coupling
is  $\sim 10^{5}$ times smaller, as noted in \S\ref{sec:HVBKT1}.
We observe that the normal fluid exhibits a similar number of
vortical structures, which are better developed and more regular
than in Figure \ref{fig:fig1}a, and
the superfluid flow pattern
is nearly unchanged from Figure \ref{fig:fig1}b.
Meridional circulation is apparent in all the results; we
find axial velocity components in the range
$10^{-5} \lsim (\vn)_{z}, (\vs)_{z} \lsim 10^{-2}$.

The flow pattern in Figure \ref{fig:fig1} changes with temperature
just like in cylindrical Couette flow.
The superfluid behaves like a classical, viscous fluid near the critical 
temperature. The Taylor vortices elongate axially, 
with a more complex pattern of eddies and counter-eddies emerging,
as the temperature is lowered \citep{hbj95,hb95}. 
For $\Rey \lsim 268$,
anomalous modes emerge in the normal component, as in
classical axisymmetric flow \citep{lm85},
characterized by Ekman cells rotating in the opposite
sense to the classical flow \citep{hb00} except
near the critical temperature.
For $\Rey \gsim 268$, the tension becomes less important than the
mutual friction and the superfluid component comes to
resemble the normal component \citep{hb95}.

\section{Glitch-induced turbulent-laminar transition}
\label{sec:VTMODEL}
In this section, we investigate how the vorticity state
of the outer core changes before,
during, and after a glitch, in the context of a generic
glitch scenario where the trigger mechanism of the glitch 
is unspecified.

\subsection{Before a glitch: vortex tangle}
Laboratory experiments on
the attenuation of second sound in narrow channels \citep{vinen57c, sbd83},
and numerical simulations based on the vortex filament method
\citep{tab03}, show that, in a rotating container of He II,
an axial (along $z$) counterflow  
$v_{ns} > v_{{\rm DG}} = 2(2 \Omega_2 \nu_s)^{1/2}$
excites growing Kelvin waves which
destabilize a rectilinear vortex array \citep{gjo74}.
The Kelvin waves grow exponentially until reconnection between
adjacent vortices occurs, generating a dense tangle.\footnote{The Kelvin waves
excited by the DGI
are unrelated to the Kelvin waves generated
by oscillations of the nuclear lattice in the inner crust \citep{eb92}.}
For typical neutron star parameters, one finds
\begin{equation}
\label{eq:dgi}
\vdg = 1.56 \left( \frac{\Omega_*}{10^2 \, \radsec}\right)^{1/2} \, \cmsec, 
\end{equation}
taking $\ln (b_0/a_0) = 20$.

The axial component of the Ekman counterflow
in the outer core of a typical neutron star generally exceeds
the instability threshold (\ref{eq:dgi}).
This is illustrated by Figure \ref{fig:fig2}, a greyscale
plot of $(\vv_{ns})_z/\vdg$; the DGI is active in regions 
with $(\vv_{ns})_z/v_{\rm DG} \geq 1$.
Figure \ref{fig:fig2}a corresponds to
HV mutual friction (initially rectilinear vortex array); 
Figure \ref{fig:fig2}b corresponds to GM mutual friction
(after a tangle forms).
In both figures, the DGI is active in most of the computational
domain, implying that {\it an initially rectilinear array
is disrupted and, once disrupted, stays that way}. 
This result is extracted empirically from the simulations.
The inclusion
of compressibility and hence stratification \citep{ae96} restricts
the DGI region to a thin boundary layer,
but the overall conclusion is the same (see \S\ref{sec:strat}).

The axial counterflow that excites the DGI is driven by Ekman pumping. 
As the crust spins down electromagnetically at
a rate $\dot{\Omega}_*$, differential
rotation builds up between the crust and outer core, with 
\begin{equation}
\Delta \Omega_{{\rm em}} =  3.16 \times 10^{-6} 
\left(\frac{\dot{\Omega}_*}{10^{-13} \, {\rm rad} \, {\rm s}^{-2}}\right)
\left( \frac{t}{1 \, {\rm yr}}\right) \, {\rm rad} \, {\rm s}^{-1},
\end{equation}
where $t$ is the time elapsed since the last glitch \citep{lg06,lss00}. 
The differential rotation induces meridional circulation \citep{green68,r93}: 
an Ekman boundary layer, with  $(\vn)_\theta \sim R_* \Delta \Omega$,
develops on a time-scale $\sim 2 \pi/\Omega_*$, and grows radially to 
a thickness
$\dEkman \approx \Rey^{-1/2} R_*$ {\rm cm} on a time scale
$t_{\rm E} \approx \Rey^{1/2} (\Omega_*)^{-1}$, 
spinning up the interior fluid \citep{a03,acg05}. 
Only the normal fluid is Ekman
pumped directly; the superfluid is spun up by the normal fluid through the mutual
friction force \citep{acg85,r93}. 

In order to estimate the counterflow velocity,
we compute $(\vns)_z$ empirically from numerical experiments and scale up
the results to neutron star parameters. For example, in a typical
set of runs with 
$\delta=0.3$ (HV mutual friction), $0.4$ (HV mutual friction),
and $0.5$ (GM mutual friction), and $\Rey=3 \times 10^4$,
we find $(\vn)_z \approx 3.15 (\vs)_z$, $(\vn)_z \approx 0.78 (\vs)_z$ and $(\vn)_z \approx 0.06 (\vs)_z$, respectively, on average over the grid,
implying $(\vns)_z \sim (\vn)_z \sim R_* \Delta \Omega_{\rm em}$, as
a general rule and hence
\begin{equation}
\label{eq:vns_ep}
(\vns)_z = 
3.16 \left( \frac{\dot{\Omega}_*}{10^{-13} \, {\rm rad} \,{\rm s}^{-2}} \right)
\left( \frac{t}{1 \, {\rm yr}} \right) \, {\rm cm} \, {\rm s}^{-1}
\end{equation}

In a typical neutron star, the Reynolds number in the outer core 
(temperature $T$) is very large, viz.

\begin{eqnarray}
\Rey& =&  1.67 \times 10^{8} \left(\frac{\rho_n}{10^{15} \, {\rm g} \, {\rm cm}^{-3}}\right)^{-1}  
\left( \frac{T}{10^8 \, {\rm K}}\right)^2 \nonumber \\
\label{eq:rey}
& &  \times \left(\frac{\Omega_*}{10^2 \, \radsec}\right),
\end{eqnarray}
given the standard viscosity $\nu_n$ resulting from
electron-electron scattering  \citep{fi79,cl87,acg05}. Consequently, the flow
is likely to be turbulent \citep{alpar78}. Our numerical simulations
cannot access this regime due to computational limitations \citep{pmgo05a}. However, we know
from laboratory experiments with He II that the flow
at $\Rey \gsim 10^{5}$ closely resembles
classical, Navier-Stokes turbulence: the superfluid and
normal fluid vorticity evolve in concert, because
the vortex lines are locked to the normal fluid eddies by mutual friction 
\citep{an75,bsbd97}. 
\footnote{In He II experiments, it is possible to reach a vortex density
$\sim 10^{7}$ {\rm cm}$^{-2}$ \citep{snd00}, comparable
to that in a neutron star, viz. $10^{5} (\Omega_*/10^{2}\, \radsec)$ {\rm cm}$^{-2}$
\citep{bel92,lg06}.} 

For the DGI to be triggered in superfluid turbulence, two conditions
must be met. First, there must be vortex segments directed parallel to the
counterflow, i.e. $\mbox{\boldmath$\omega$}_s \cdot \vv_{ns} \neq 0$ \citep{bsbd97}. As the 
superfluid is locked to the normal fluid, and the vortex density is high, the
condition $\mbox{\boldmath$\omega$}_s \cdot \vv_{ns} \neq 0$ is satisfied locally throughout the core.
Second, a vortex segment is locally unstable to (helical) Kelvin wave
perturbations of wavelength $\lambda$ if the local counterflow exceeds
the critical speed $v_{\rm DG}^{\rm turb} = \kappa \ln(\lambda/2 \pi a_0) /(2\lambda)$
\citep{bsbd97}. Large wavelengths (of the order of the size of the star)
are most unstable, with $\lambda \sim 10^{6}$ {\rm cm} giving
$v_{\rm DG}^{\rm turb} = 3 \times 10^{-2}$ {\rm cm s}$^{-1}$. In this regard, 
the DGI 
threshold (\ref{eq:dgi}) is conservative ($v_{\rm DG}^{\rm turb} < \vdg$).
\footnote{A vacillating flow \citep{btma04} can also trigger vortex oscillations
by inducing an oscillating counterflow. A perturbation
with wavelength $\lambda$ grows if the vacillation frequency $\omega_0$ is
mantained over a time-scale
$t \leq \lambda v_{ns}/2 \pi \nu_s  \omega_0$ \citep{btma04}.}

\subsection{Glitch: decay of the tangle}
During a glitch event, the differential rotation instantaneously
shuts off and the tangle
starts to decay.
In a self-sustaining vortex tangle,
Kelvin waves grow until their amplitude approaches 
the average vortex separation, whereupon the vortex lines continuously 
reconnect to form loops \citep{jm04}. Reconnection stops once the counterflow
ceases, as happens immediately after a glitch, when the outer core
and inner crust come into corotation ($\Omega_1 = \Omega_2$). The decay
time-scale $\tau_{\rm d}$ equals, to a good approximation, the 
reconnection time-scale just before the counterflow
ceases. 

To estimate $\tau_{\rm d}$, we approximate
the vortex loops by rings of radius $R$, whose
characteristic lifetime is given by
$\tau_{\rm d} = R^2(2\nu_s \alpha)^{-1}$, with $\alpha = B \rho_n/2 \rho$
\citep{bdv83,tbam04}.
The radius of a ring can be estimated from
the vortex line density $L$ (length per unit
volume), with $R = 0.5 L^{-1/2}$ in a steady-state tangle.
The evolution of $L$ is
given by Vinen's equation \citep{vinen57c}, generalized
by \citet{jm04} to include rotation:
$$ 
\frac{1}{\Omega_2} \frac{dX}{dt} = - \alpha_{3} X^{2} +
\left[ \frac{\alpha_1 v_{ns}}{(\kappa \Omega_2)^{1/2}} 
+ \beta_{2} \right] X^{3/2}
$$
\begin{equation}
\label{eq:vinen_jou}
- \left[ \beta_1 + \frac{\beta_4 v_{ns}}{(\kappa \Omega_2)^{1/2}}\right] X,
\end{equation}
with $X = \kappa L /\Omega_2$, where $\alpha_i$, $\beta_i$
are dimensionless
friction coefficients, whose values depend
on the temperature \citep{mj05}. 
By fitting to experimental data,
one finds $\alpha_3 = 20.0 \alpha_1$, $\beta_1 = 35.6 \alpha_1$, $\beta_2 = 53.6 \alpha_1$,
and $\beta_4=1.43 \alpha_1$ \citep{sbd83,jm04}.
From experiments, we also have  $\alpha = \chi_1 \alpha_1$
and $\chi_1 \sim 1$ \citep{vinen57c}, so we take $\alpha = \alpha_1$ below.
Setting $dX/dt = 0$ in (\ref{eq:vinen_jou}) and solving 
for $L$ in the steady state, we find that the 
tangle decays on a time-scale

\begin{eqnarray}
\label{eq:decay_tangle1}
\tau_{\rm d} &  =& \frac{2.5 \times 10^3 \kappa}{\alpha v_{ns}^2 \ln(b_0/a_0)} \\
\label{eq:decay_tangle2}
& \approx & 7.6 \times 10^{5} \left(\frac{\alpha}{10^{-7}}\right)^{-1} \nonumber \\
& &  \times \left(\frac{\dot{\Omega}_*   }{10^{-13} \, {\rm rad} \, {\rm s}^{-2}}\right)^{-2}
\left( \frac{t}{1 \, {\rm yr}} \right)^{-2} {\rm s},
\end{eqnarray}
where $v_{ns}$ is the counterflow speed immediately before
the differential rotation shuts off, given by (\ref{eq:vns_ep}).

Estimates for the friction parameter $\alpha$ are uncertain
\citep{sonin87}, but calculations based on electron scattering
by proton vortex clusters and magnetized vortex cores
give $B \approx 10^{-4}$ \citep{sss82,sc98}.
\footnote{
An absolute lower limit is $B = 10^{-18}$, calculated
from electron
scattering off vortex clusters in the high-density regime,
where the relaxation time-scale equals the age of the pulsar
[\citet{sc98}; A. Sedrakian 2006, private communication].}
This implies $\alpha \lsim 10^{-7}$
for $\rho/\rho_n = 10^{-2}$ in the outer core \citep{ss95}.
Therefore $\tau_{\rm d}$ is greater than the glitch trigger
time-scale, as assumed a priori. Note that the 
$B$ and $B^\prime$ parameters
were also calculated by \citet{m91a} and \citet{asc06}, who
found $B^\prime = B^2$ and
\begin{eqnarray}
\label{eq:bbprime1}
B & \approx&  4 \times 10^{-4} \left( \frac{m_p- m_p^*}{m_p} \right)^2
 \left( \frac{m_p}{m_p^*} \right)^{1/2}
\left( \frac{x_p}{0.05}\right)^{7/6} \nonumber \\
& & \times \left( \frac{\rho}{10^{14} \, {\rm g} {\rm cm}^{-3}}\right)^{1/6}
\end{eqnarray}
where $x_p = \rho/\rho_p$ is the proton fraction, and
$m_p$ and $m^*_p$ are the bare and effective masses
of the proton respectively \citep{asc06}. Equation
(\ref{eq:bbprime1}) includes modifications from
entrainment \citep{m91a,asc06}. In the outer
core, with $\rho = 2.8 \times 10^{14}$ {\rm g} {\rm cm}$^{-3}$
and $x_p = 0.038$ \citep{m91a}, we obtain $B=2.7 \times 10^{-4}$,
approximately the value quoted by \citet{sss82}
and \citet{sc98}.

In laboratory and numerical experiments, it is observed that
the vortex tangle is polarized; 
the average vorticity projected along the
rotation axis is not zero \citep{finne03,tbam04,tk05}. This is because, under
these experimental conditions, one has
$\beta_1 \gsim \beta_4 v_{ns}/(\kappa \Omega_2)^{1/2}$ 
in equation (\ref{eq:vinen_jou}), where $\beta_1$
is the polarization-inducing term. Naively, one might
expect the tangle to be even more polarized in a rapidly
rotating neutron star, but in fact the opposite is true: the
tangle is less polarized, because we have
$\beta_1 \ll \beta_4 v_{ns}/(\kappa \Omega_2)^{1/2}$
from Figure \ref{fig:fig2}. A thorough study of this issue,
including whether the large-$\Omega_2$ limit
considered by \citet{jm04}
in deriving equation (\ref{eq:vinen_jou}) is applicable
for $v_{ns} \gg (\kappa \Omega_2)^{1/2}$, lies
outside the scope of this paper.

\subsection{After a glitch: reformation of the tangle}

After the tangle decays to a rectilinear array, differential rotation 
due to electromagnetic spin down
between the inner and outer shells begins to grow until $(\vns)_z$ 
exceeds $\vdg$. From equations (\ref{eq:dgi}) and (\ref{eq:vns_ep}), this
occurs after a time 

\begin{equation}
\label{eq:reform} t_{\rm tan} = 0.49
\left( \frac{\Omega_*}{10^2 \, {\rm rad} \, {\rm s}^{-1}} \right)^{1/2} 
\left( \frac{\dot{\Omega}_*}{10^{-13} \, {\rm rad} \, {\rm s}^{-2}} \right)^{-1} \, {\rm yr},
\end{equation}
whereupon a vortex tangle develops again in the outer core, via the DGI.
This triggers a switch from HV to GM mutual friction,
weakening the coupling between the normal
fluid and superfluid components.

The characteristic time-scale for the tangle to develop 
equals the growth rate of helical vortex perturbations
(Kelvin waves) via the DGI. By linearizing
Schwarz's equation \citep{s85,s88}, in the vortex filament model, 
one finds that the
fastest growth occurs at a wavenumber $k = v_{ns}/2\nu_s$, with
\citep{tbam04}
\begin{eqnarray}
\label{eq:growth_tangle1}
\tau_{\rm g} &  = & \frac{ \kappa \ln(b_0/a_0)}{\pi \alpha v_{ns}^2} \\
\label{eq:growth_tangle2}
& \approx & 4.9 \times 10^{4} \left(\frac{\alpha}{10^{-7}}\right)^{-1}
\left(\frac{\Omega_*}{10^2 \, {\rm rad} \, {\rm s}^{-1}}\right)^{-1} {\rm s},
\end{eqnarray}
where (\ref{eq:growth_tangle2}) follows from (\ref{eq:growth_tangle1})
by substituting (\ref{eq:dgi}). Hence, for 
typical neutron star parameters, the tangle is reestablished over
$\sim 1$ {\rm d} after $\sim 1$ {\rm yr} elapses following
a glitch, less than the inter-glitch interval observed in
most pulsars \citep{sl96,lss00}. Note that $\tau_{\rm g}$ does not 
equal $\tau_{\rm d}$; the time required for a tangle to grow from
a rectilinear array is shorter than the time required for an
existing tangle to decay back to a rectilinear array, provided
that $t < 8.0 t_{\rm tan}$ (and longer otherwise). Note also that
$v_{ns}$ just before the glitch ($\sim R_* \Delta \Omega$),
which appears in (\ref{eq:decay_tangle1}), is typically
greater than $v_{ns}$ at $t_{\rm tan}$ ($\sim \vdg$), which appears
in (\ref{eq:growth_tangle1}), provided that $t > t_{\rm tan}$.

Equation (\ref{eq:growth_tangle2}) gives the minimum time for a
tangle to reform, assuming it does so simultaneously everywhere
in the outer core of the star. In practice, $v_{ns}$ does not
exceed $v_{\rm DG}$ everywhere simultaneously. Regions where the DGI
is activated are interspersed with regions where it is not, even
at $r=R_1$ and $R_2$, so that the transition from HV to GM friction
is more gradual than equation (\ref{eq:growth_tangle2}) suggests.
This important issue, which is also relevant to the interpretation
of the laboratory experiments performed by \citet{tsatsa80},
is explored briefly in \S\ref{sec:summary}. However, a thorough
numerical study lies outside the scope of this paper.

\section{Rotational evolution of the crust}
\label{sec:rotevol}

In this section, we calculate numerically the torque 
acting on the stellar crust during a 
transition from a state of turbulent vorticity (tangle)
to laminar vorticity (rectilinear array), by solving the hydrodynamic 
HVBK equations 
in a differentially rotating spherical shell. Global simulations
of this type are necessary to calculate the
observable response of a neutron star, e.g. $\Omega(t)$ and $\dot{\Omega}(t)$, to
the internal physics elaborated in \S\ref{sec:VTMODEL}.

\subsection{Numerical method and parameters}

We solve equations (\ref{eq:supeq1}) and
(\ref{eq:supeq2}) using a pseudospectral collocation
method to discretize the problem \citep{boyd02,canuto88} and 
a time-split algorithm to step forward in time \citep{canuto88},
as described by \citet{pmgo05a}.
The velocity fields are
expanded in a restricted Fourier series in $\theta$ and
$\phi$ and a Chebyshev series in $r$ \citep{orszag74,boyd02},
with $(N_r,N_\theta,N_\phi) = (120,250,4)$  modes required
to fully resolve the flow. The initial time-step $\Delta t = 10^{-4}$
is lowered to $\Delta t=10^{-5}$ during the GM $\rightarrow$ HV
transition to prevent spurious oscillations in the torque
\citep{pmgo05a}. Instabilities arising from
the sensitivity of spectral methods to the boundary
conditions \citep{pmgo05a}, and oscillations due to the
Gibbs phenomenon \citep{gho84,canuto88}, are smoothed using
a low-pass spectral filter \citep{don94}, a common practice
\citep{osher84,mittal99}.

We adopt parameters as close to those of a realistic neutron star
as our computational resources permit.
In the outer core, we take
$\rho_s/\rho = 0.99$ and $\rho_n/\rho=0.01$ \citep{bel92},
and hence $B = 1.5$, $B^{\prime} = 0.9$, 
and $A^{\prime} = 1.0 \times 10^{-4}$ at the corresponding scaled
temperature $T/T_c$ in He II \citep{bdv83,db98}. These friction
coefficients are $\sim 10^4$ times greater than those used
in the analytic estimates in Section \ref{sec:VTMODEL}. We
adopt the higher values deliberately. Otherwise, the effects
introduced by $B$ and $B^\prime$ would take too long to
build up; computational limitations prevent us from
simulating more than $\sim 10$ rotation periods.
(For the same reason, our $\Delta \Omega/\Omega$
is unrealistically large.)
Realistic
Reynolds numbers, estimated from equation (\ref{eq:rey}), 
are too high to be simulated directly, so we restrict ourselves
to the range $10^{3} \lsim \Rey \lsim 10^{5}$; the most
stable evolution occurs  for $\Rey \approx 3 \times 10^{4}$, corresponding
to an Ekman number $(\Rey)^{-1} \approx 3 \times 10^{-5}$ that is
two orders of magnitude smaller than in a typical neutron star \citep{ae96}.
The tension force is dominated by mutual friction,
except in relatively old ($> 10^{4}$ {\rm yr})
neutron stars \citep{g70}, with $\nu_s \sim 10^{-18}$ in our dimensionless units.
For $10\leq A \leq 10^4$, we find empirically that the torque
depends weakly on the GM force, so we use $ 10 \leq A \leq 10^{2}$ 
to obtain identical results at lower computational cost.

\subsection{Turbulent-laminar transition}
\label{sec:turlam_num}
We consider a turbulent initial state 
with $\Omega_1 > \Omega_2$ and $(\vv_{ns})_z > \vdg$ everywhere, such that 
the normal and superfluid
components are coupled by GM mutual friction.
The transition from a vortex tangle to a rectilinear vortex array is
simulated by changing the mutual friction suddenly,
from GM to HV, and simultaneously spinning up
the outer shell, such that $\Omega_2 = \Omega_1$.
This occurs at $t=20$ in all the figures to follow.
The angular shear before the transition lies in the range
$0.1 \leq \Delta \Omega \leq 0.3$, for 
gaps in the range $ 0.2 \leq \delta \leq 0.5$. 
Note that we are forced to choose $\Delta \Omega$ unrealistically large,
compared to typical neutron star values, in order to allow the
inner and outer surfaces of the shell to
``lap" each other within a reasonable run time.\footnote{Flow structures
associated with differential rotation develop on the time-scale $(\Delta \Omega)^{-1}$.}
Various combinations of  $\Delta \Omega$ and  $\delta$ affect
the viscous torque
on the inner and outer surfaces during steady 
differential rotation in different ways, as discussed below.

Streamline snapshots for a run with $\delta = 0.5$ and $\Delta \Omega = 0.2$
are presented in Figures \ref{fig:fig3}a--\ref{fig:fig3}f,
with the turbulent-laminar transition 
GM $\rightarrow$ HV, $\Omega_2 \rightarrow \Omega_1$ 
occurring at $t=20$. 
The HV mutual friction, with 
$|\efe_{\rm HV}|/|\efe_{\rm GM}| \sim 10^{5}$, 
couples the normal and superfluid components strongly. 
\footnote{The empirical result $|\efe_{\rm HV}|/|\efe_{\rm GM}| \sim 10^{5}$
is supported by order-of-magnitude estimates \citep{pmgo05a}
but should be treated with caution, because the microphysics
of the GM friction force (and hence the value of $A^\prime$)
has not yet been worked out rigorously in a neutron
superfluid.}
Both components exhibit a quasi-periodic
motion, with a secondary circulation cell filling
most of the shell, and 
two or three smaller vortices emerging intermittently.
An oscillatory polar Ekman cell also exists initially,
disappearing at $t=140$. 

We can compare this behaviour to
the flow without a turbulent-laminar transition.
Figures \ref{fig:fig4}a--\ref{fig:fig4}f display a sequence of
streamline snapshots for $\delta=0.5$ and $\Delta \Omega=0.2$, with GM
mutual friction at all times. 
At $t=18$, three polar cells are observed in
both components, together with a 
secondary cell at mid-latitudes.
The latter cell elongates after $t=20$ and
is present only in the superfluid component at
later times. At the equator, the flow
switches between one and three cells, but
this behavior persists after $t=100$ only in
the superfluid component, while the normal
fluid decouples increasingly.

The axial counterflow is
significant at all times in Figures \ref{fig:fig4} and \ref{fig:fig5}.
Scaling to neutron star parameters, we find
$v_{ns}/v_{\rm DG} \sim 10^{3} (\Delta \Omega / 10^{-4} \, \radsec)
(\Omega_*/10^2 \, \radsec)$ near the equator,
and  
$v_{ns}/v_{\rm DG} \sim 10^{6} (\Delta \Omega / 10^{-4} \, \radsec)
(\Omega_*/10^2 \, \radsec)$ near the poles, at $t=200$.

Figures \ref{fig:fig5}a--\ref{fig:fig5}d display $\Delta \Omega / \Omega$ 
as a function of time after the turbulent-laminar transition
for $\delta=0.2$, $0.3$, $0.4$, and $0.5$, respectively.
\footnote{Note that $\Delta \Omega / \Omega_*$
is not calculated self-consistently:
it is obtained by numerically integrating the viscous
torque on the outer sphere, after solving (\ref{eq:supeq1})
and (\ref{eq:supeq2}) subject to the boundary
condition $\Omega_2(t)={\rm constant}$, i.e. we do not include
the (slight) evolution of $\Omega_2(t)$ when computing the fluid flow.}
The evolution is qualitatively similar
in all the cases considered.
There is an initial transient, in which the flow adjusts to
the initial spin up on a time-scale $t \sim 6$,
just like in the classical (viscous)
spin-up problem \citep{green68}.
The fractional torque deviation,
$\Delta \dot{\Omega} / \dot{\Omega}$,
is plotted as a function of time in Figures \ref{fig:fig6}a--\ref{fig:fig6}d. 
An initial jump, coinciding with the sudden spin up,
is observed
in all the panels, followed by a rapid exponential
decay and persistent, small-amplitude oscillations. 
We find an $e^{-1}$ relaxation time of $\tau \sim 2$, $1.6$, and $1.6$ for
$\Delta \Omega = 0.1$, $0.2$, and $0.3$ respectively, with
$\tau$ essentially independent of $\delta$.
In contrast, in the inverse experiment, where
the mutual friction changes from HV to GM, \citet{pmgo05a}
observed that
$\Delta \dot{\Omega}$ is
almost constant after the post-glitch transient.
The period of the oscillations does not
change with $t$ for $t \gsim 100$.

\subsection{Effect of the superfluid}

The effect of the superfluid on the evolution can be 
analyzed with the help of Figure \ref{fig:fig8}a, which
plots the torque on the outer sphere
as a function of time (solid curve) in the absence
of spin up ($\delta=0.5$, $\Rey=3 \times 10^{4}$,
$\Delta \Omega = 0.1$, GM friction). For comparison, the
dashed curve plots the
torque for a classical
viscous fluid with the same parameters mentioned
above. The evolution is qualitatively similar in both cases,
displaying oscillations with the
same periodicity and similar amplitude, but the
curves diverge for $t \gsim 20$ as the viscous
fluid evolves more rapidly to a stationary
state, driven by classical Ekman pumping.
Classical theory
predicts a time-scale $t_{\rm E} \approx 173$ in the limit
where the Rossby number $\Delta \Omega/\Omega$ vanishes;
an exponential fit to the dashed curve gives $t_{\rm E} \approx 243$,
which agrees well given that we have $\Delta \Omega/\Omega = 0.1$
in our numerical experiments.
In a superfluid, on the other hand, the torque decays faster initially,
with time constant
$\sim 50$, then starts to increase again for
$t \sim 240$, as part of a long-period oscillation whose ultimate
fate is unknown as it occurs over a time-scale beyond our computational
limit.

Figure \ref{fig:fig8}b shows
the evolution of the fractional change in angular
velocity after spin up ($\Omega_2 = 0.9 \rightarrow 1.0$ at
$t=20$) in three scenarios: instantaneous
change in mutual friction, GM $\rightarrow$ HV (solid curve);
constant GM friction (dashed curve);
and a classical viscous fluid (dashed dotted curve).
The classical viscous fluid and constant
GM friction evolve similarly.
The torque decays exponentially on a time-scale $\sim 2.15$, then
decays almost linearly with $t$. By contrast, in
the GM $\rightarrow$ HV transition, the torque decays
exponentially, on a time-scale $\sim 2.20$, then
oscillates with peak-to-peak amplitude $\approx 0.4$ 
(in units of $\rho R_2^5 \Omega_1^{2}$)
and period $\sim 6$. In other words, the oscillations are
sustained by the HV friction, which is much weaker than
its GM counterpart. Note that the oscillations
do not correspond to Tkachenko waves, since the HVBK equations
assume that the free energy of the vortex array depends
only on the vortex line density, not on vortex
lattice deformations \citep{cb83,cb86,donnelly91}.

\subsection{Streamlines}
The meridional streamlines
for the transitions described in the previous paragraph
evolve as in Figure \ref{fig:figc}.
Figures \ref{fig:figc}a--\ref{fig:figc}b show the streamlines
for a classic viscous fluid before ($t=20$) and 
after ($t=22$) the glitch. There
is a rapid redistribution of vorticity near the outer sphere, 
with two circulation cells replaced
by one elongated meridional shell, while the flow
near the inner sphere hardly changes, with 
one circulation cell near the equator
and one near the pole. Figures \ref{fig:figc}c--\ref{fig:figc}d
show the meridional streamlines for the normal
component of the superfluid.
The flow resembles a classical viscous fluid
(cf. Figure \ref{fig:figc}a), although it does become
more complex after the spin up.
As shown in Figure \ref{fig:figc}d,
the large cell near the outer
sphere is replaced by one primary circulation cell
and three secondary cells (including one near the
poles).
Note that the greatest contributions
to the torque come from the mid-latitude
regions where the structure of eddies
and counter-eddies is richer, as in
Figure \ref{fig:figc}d. Finally, in Figures \ref{fig:figc}e--\ref{fig:figc}f,
we see the meridional streamlines for the inviscid component
of the superfluid.
The pattern before the jump
(Figure \ref{fig:figc}e) differs from the normal fluid; the 
GM friction creates additional
eddies near the outer sphere and closer
to the poles. However, after the transition to HV friction,
the inviscid component comes to resemble the normal component,
as can be seen by comparing Figure \ref{fig:figc}f with
Figure \ref{fig:figc}d. This occurs because
the two components are coupled more strongly by 
HV friction
($|\efe_{\rm HV}|/|\efe_{\rm GM}| \sim 10^{5}$),
so that the superfluid is dragged along by the normal
fluid.

\subsection{Stratification}
\label{sec:strat}

Gravitational stratification can strongly suppress
Ekman pumping, as
the Ekman layer is squashed close to the outer sphere
\citep{cctl71,ae96}.
Nevertheless, the Ekman
layer, however thin, always exists, in order to satisfy the boundary conditions
at $r=R_2$. It contains a meridional counterflow
given by equation (\ref{eq:vns_ep}).
Consequently, the transition from turbulent to laminar vorticity
(and vice versa), which relies on such a meridional counterflow,
still occurs in a stratified star, and its dramatic
effect on the torque is the same. The
Ekman layer, no matter how thin, acts like a film of ``oil"
between two sliding surfaces (here, the outer core
and inner crust), whose coupling strength (``stickiness")
changes abruptly when the meridional speed of the
``oil" exceeds a threshold. In other words, 
equations (\ref{eq:vns_ep}), (\ref{eq:decay_tangle2}),
(\ref{eq:reform}), and (\ref{eq:growth_tangle2}), and
the conclusions that follow from them, are unaltered
by stratification; even though the volume of the outer
core occupied by the vortex tangle (which does
not affect the torque on the crust directly) does change.

The Ekman layer thickness
decreases as ${\rm e}^{-\kappa_Y} \dEkman$, where
$\dEkman$ is the thickness without stratification
and $\kappa_Y$ is the compressibility of the fluid.
The Ekman time-scale increases
as $({\rm e}^{\kappa_{Y}}-1)\kappa_Y^{-1} t_{\rm E}$, where
$t_{\rm E}$ is the time-scale without stratification.
The fluid is restricted to move on
concentric spherical shells \citep{ld04}.
Importantly, $({\rm e}^{\kappa_{Y}}-1)\kappa_Y^{-1} t_{\rm E}$
is the time for the Ekman layer to
extend throughout the outer core, not the
time required to establish the meridional flow at
$r \approx R_2$ ($\sim \Omega_1^{-1}$),
which controls the onset of the DGI. 

It is conceivable
that the DGI is
excited in shells at certain radii
where $v_{ns}$ peaks, so that we end up with
a sequence of shells containing alternating laminar and turbulent 
superfluid vorticity. In this scenario, the detailed study
of which lies outside the scope of this paper,
the stability of the vorticity configuration is restricted
by the Richardson criterion, which states that the
configuration is Kelvin-Helmholtz unstable for
$ N^2|\partial v_\phi/\partial r|^{-2} < 1/4$ \citep{mm05}.
In a neutron star, the Brunt-V\"ais\"ala frequency
is $N \approx 5 \times 10^{2} \, {\rm s}^{-1}$ \citep{rg92}. 

The effects of stratification are not considered in detail
in this paper because they cannot be studied properly with our
numerical method; the pressure projection step only works
with divergence-free velocity fields \citep{bb02,pmgo05a}.
However, a crude approach to get a feel for the effects is
to numerically suppress $v_r$ using
a low-pass exponential filter \citep{don94}, viz.
$v_r \rightarrow \exp[-
(k/N_{r})^{\gamma} \ln \epsilon ] v_r $, with $0
\leq |k| \leq N_{r}$, where
$\epsilon =  2.2 \times 10^{-16}$ is the machine zero,
and $\gamma$ is the order of the filter. If we
ramp up $\gamma$ with $r$ as 
$\gamma =  r(n_2-n_1)/(R_2-R_1)+n_1-R_1(n_2-n_1)/(R_2-R_1)$,
with $n_1=2$ and $n_2=12$,
then $v_r$ is weakly suppressed near $R_2$
($\gamma=12$) and
strongly suppressed near $R_1$ ($\gamma=2$), so
that the flow is confined approximately to concentric shells.
Preliminary results, for a viscous Navier-Stokes fluid,
show that the filtering reduces the meridional
circulation, flattening the streamlines radially.


\section{Summary and discussion}
\label{sec:summary}

In this paper, we investigate how transitions between turbulent and laminar
states of superfluid vorticity alter the standard theoretical picture
of pulsar rotational irregularities like glitches and timing noise.
(i) Most of the time, except in the immediate aftermath of a glitch, differential
rotation in the outer core drives a nonzero, poloidal
counterflow which continuously excites the DGI. A vortex tangle is thereby
maintained in the outer core. The mutual friction in this regime, which is
of GM form, couples the normal and superfluid components loosely. (ii) Immediately
after a glitch, the differential rotation ceases, as does the poloidal
counterflow. The vortex tangle decays over the mean life-time of
its constituent vortex rings, $\tau_{\rm d} = 
7.6 \times 10^{5} (\alpha/10^{-7})^{-1} (\dot{\Omega}_*/10^{-13} \, {\rm rad} \, {\rm s}^{-2})^{-2} ( t/1 \, {\rm yr})^{-2}$ {\rm s}.
A rectilinear vortex array develops, and the mutual friction switches
to HV form, coupling the normal and superfluid components much more strongly.
(iii) After $t_{\rm tan} = 0.49 \, (\Omega_*/10^2 \, {\rm rad} \, {\rm s}^{-1})^{1/2}(\dot{\Omega}_*/10^{-13}
\, {\rm rad} \, {\rm s}^{-2})^{-1}$ {\rm yr}, electromagnetic spin down
builds up the differential rotation sufficiently to
drive a poloidal counterflow that exceeds the DGI threshold.
A vortex tangle forms again in a time 
$\tau_{\rm g} = 4.9 \times 10^{4} (\alpha/10^{-7})^{-1}(\dot{\Omega}_*/10^{-13} \, {\rm rad} \, {\rm s}^{-2})^{-1}$ {\rm s},
and the mutual friction reverts to GM form. Note that vortex pinning 
provides the 
boundary conditions for the superfluid
SCF but does not occur within the outer core itself \citep{dp03}.
{\it Therefore our new phenomenological picture is not a complete model
for glitches}. It merely clarifies the vorticity state of the outer
core before and after a glitch as an input into future models than
incorporate the full glitch dynamics, including trigger mechanisms
related to pinning in the inner crust.

We draw together the strands
of the model in Figure \ref{fig:fig9}, which displays the
evolution of the torque and the regions where the 
DGI is active during the following numerical experiment:
we fix $\Delta \Omega = 0.1$ until
$t=20$, accelerate the outer sphere instantaneously to corotation
at $t=20$, then decelerate
the outer sphere according to $\Omega_2(t) = 1 -
0.001(t - 20)$ for $t>20$. This mimics the situation in
a real pulsar, where we have $t_{\rm E} \ll t_{\rm tan}$,
i.e. Ekman pumping brings the fluid into corotation {\it before}
the DGI gradually starts being reexcited
throughout the outer core. To make the experiment
as realistic as possible, we do not
assume that the mutual friction takes the same form
everywhere in the outer core, but rather
choose GM or HV friction at each point according to whether
$(\vv_{ns})_z$ is greater or less than $v_{\rm DG}$ locally.
In this comparison, we approximate $\vv_{ns}$ by $\vv_n$,
as in equation (\ref{eq:vns_ep}),
because $\vv_s$
can become very complicated (e.g. Figure \ref{fig:fig1}),
creating numerical difficulties.
In order to satisfy $t_{\rm E} \ll t_{\rm tan}$
while keeping $\Delta \Omega$ large enough
so that the spheres ``lap" each other several times,
we are forced by computational exigencies to adopt
a relatively low Reynolds number $\Rey = 100$,
shortening the Ekman time ($t_{\rm E} \sim 10 \Omega_1^{-1}
\ll t_{\rm tan}$), and to
artificially boost $v_{\rm DG}$, so that
$|(\vv_{n})_z/v_{\rm DG}|$ does not exceed $\sim 3$ throughout
the computational domain.

The results of the above numerical experiment are presented
in Figure \ref{fig:fig9}.
Contours of $|(\vv_{n})_z/v_{\rm DG}|$, before and after the spin up at $t=20$,
are plotted in Figures \ref{fig:fig9}a--\ref{fig:fig9}e. Shaded
regions indicate where the DGI is active, i.e. 
$|(\vv_{n})_z/v_{\rm DG}| > 1$.
Just before the glitch (Figure \ref{fig:fig9}a), $32$ \% of the superfluid
is in a turbulent state, with the DGI active
close to the inner sphere and at intermediate latitudes
where meridional circulation is significant.
After the differential rotation shuts off at $t=20$, the
DGI initially spreads through the shell as transient
axial flows increase, occupying
$39$ \% of the volume at $t=22$. However, the flow
quickly settles down to a state of near-corotation
during the time interval $ 24 \lsim t \lsim 50$, HV
friction dominates, and the torque decays exponentially,
with time constant $\sim 10$ (followed by 
a linear decay). At $t=50$, the DGI slowly begins
to assert itself again, starting from
the inner sphere. As for a classical viscous fluid,
when the outer sphere spins down, it pumps fluid
radially inward and along the axis of rotation \citep{vanyo},
so the axial flow speed is greatest near the inner sphere.
By $t=120$, when $\Delta \Omega = 0.1$, the vorticity state
is similar to that at $t=20$, before spin up.

One might wonder if, in a realistic neutron star, 
the superfluid ever exits the
turbulent state and becomes laminar.
For the simulations in this paper, which have
$\Rey \leq 3 \times 10^4$, the answer is yes.
Figure \ref{fig:fig9} shows that, when the glitch
occurs and the spheres come
into corotation, $|(\vv_{n})_z|$
falls below the DGI threshold after a time
$t \sim t_{\rm E}$. The vortex tangle is then
guaranteed to decay on a time-scale given
by (\ref{eq:decay_tangle1}) and (\ref{eq:decay_tangle2}),
as observed in terrestrial experiments. However, for
more realistic neutron star Reynolds numbers
($\Rey \geq 10^{8}$), which are too challenging to
simulate at present, the turbulent eddies
in the normal fluid decay more slowly when
the spheres come into corotation, so
that $|(\vv_{n})_z|$ remains above the
DGI threshold for longer. If this happens, the vortex tangle
may persist until the next glitch occurs, so
that the superfluid never exits the turbulent state.

What are the implications of Figure \ref{fig:fig9} and
the results in \S\ref{sec:VTMODEL} and \S\ref{sec:rotevol}
for observations of glitches? Before considering this
question, we emphasize again that the results in this paper
do not constitute a theory of glitches, because important
questions regarding the glitch trigger remain unresolved.
Nevertheless, some general remarks can be made. First of all,
it is clear that transitions between flow (and vorticity)
states in superfluid SCF are caused by changes in $\Rey$, and
that such transitions become more frequent and complicated
as $\Rey$ increases \citep{ybm77,je00}.
This is compatible with the observation that adolescent pulsars
($\sim 10^4$ {\rm yr} old, like Vela) glitch most actively
\citep{lss00}.
In younger pulsars (age $\lsim 10^4$ {\rm yr}), $T$ and hence $\nu_n$
are relatively high, so $\Rey$ is low. In older pulsars
(age $\gsim 10^4$ {\rm yr}), $\Omega$ and hence $\Rey$ are low
following electromagnetic spin down [although this trend
is not straightforward and can be masked by localized heating
from differential rotation between the superfluid and the
crust \citep{g75,ll99} or crust cracking \citep{lfe98,fle00}].
A systematic statistical study of glitch activity versus $\Rey$
will be published elsewhere \citep{mpw06}, but preliminary
estimates of $T$ and hence $\Rey$ from
cooling curves \citep{tsuruta74,tsuruta98,plps04} including
superfluidity \citep{fi76,fi79,acg05} give
Reynolds numbers in the range $10^{8} \lsim \Rey \lsim 10^{12}$
for glitching pulsars. Two of the most active glitchers,
the Crab and Vela, have $\Rey \sim 10^{9}$ and
$\Rey \sim 10^{10}$ respectively.
One expects that, at such high $\Rey$, the fluid is 
turbulent, with the kinetic
energy concentrated at large scales \citep{yav78,yb86},
as for a classical viscous fluid \citep{sdgv93,bsbd97}.
This suggests that superfluid turbulence in pulsar interiors
is an important factor in glitch dynamics.

If it is true that the vorticity in the outer core
exists in a turbulent state before a glitch, as
postulated in our model, then $t_{\rm tan}$ represents
a lower bound on the time between glitches. In testing
whether this bound is respected by the glitching pulsars
currently known, we are hampered by the fact that most
of these objects have only glitched once. Nevertheless, for
all the $28$ pulsars that have glitched repeteadly, we find that
the minimum inter-glitch time interval $t_{\rm min}$ is greater than
$t_{\rm tan}$, as the theory predicts \citep{mpw06}. The
object PSR $2116+1414$ approaches the bound most closely, with
$t_{\rm min}=3.9$ {\rm yr} and $t_{\rm tan} = 2.0$ {\rm yr}. This
is encouraging, because the $28$ objects cover five decades
in $t_{\rm tan}$ and three decades in $t_{\rm min}$, and the
theoretical expression (\ref{eq:reform}) for $t_{\rm tan}$
contains zero free parameters. Note that $t_{\rm tan}$ is proportional
to the characteristic age ($=\Omega_*/2\dot{\Omega}_*$) divided by
$\Omega_*^{1/2}$. Note also that the activity parameter defined
by \citet{ml90} involves glitch amplitudes (which are highly
variable) as well as mean recurrence times, so we do not
predict a correlation between the activity parameter and $t_{\rm tan}$.

It is harder to test the theoretical decay time-scale of the
vortex tangle, as predicted by (\ref{eq:decay_tangle2}), because
it remains unclear what observable features are engendered by
the decay process. The observed exponential post-glitch relaxation
is of viscous origin and occurs on a time-scale much larger
than $\tau_{\rm d}$. On the other hand, the decay of the tangle
is accompanied by a large increase in mutual friction (GM $\rightarrow$ HV), which
may be connected with the rapid jump in $\Omega$ during a glitch. The jump
in $\Omega$ has never been resolved in time, in pulsars which
are nearly constantly monitored, consistent with the predictions
of (\ref{eq:decay_tangle2}) for the Crab ($\tau_{\rm d} = 3 \times 10^{-4}$ {\rm s})
and Vela ($\tau_{\rm d} = 0.2$ {\rm s}). However, equation
(\ref{eq:decay_tangle2}) predicts that it may be possible to resolve
the $\Omega$ jump in older pulsars, provided that the time between
glitches does not increase faster than $\dot{\Omega}_*$. In making
these estimates we assume the canonical value $\alpha = 10^{-7}$ for
every object, in the absence of a microscopic theory, yet this
is clearly an oversimplification because $\alpha$ is sensitively
temperature dependent.

Oscillations in $\dot{\Omega}_*$ were observed before 
(period $\sim 10$ {\rm d}) and after (period $\sim 25$ {\rm d}) the
Vela Christmas glitch, with  $\Delta \dot{\Omega}/\dot{\Omega}_* \approx 0.17$ 
\citep{mhmk90}.
In our numerical simulations, persistent torque oscillations 
are always observed when the outer core rotates differentially, as
occurs before a glitch.
They are also observed after
a switch from GM to HV
friction, as occurs after a glitch. By comparing 
the dashed and solid curves in
Figure \ref{fig:fig8}b, we see that the oscillations
are sustained by HV mutual friction. The oscillation
period in our simulations is much shorter than in pulsar data,
because we are restricted to $\Rey \leq 3 \times 10^{4}$.
An alternative explanation is that
vortices in the inner crust oscillate relative
to the normal fluid in the core \citep{ssa95}.

Several of the effects explored in this paper have been studied
in terrestrial laboratories. Our results will motivate
new experiments of this sort, cf. \citet{alpar78} and \citet{aprs78}.
Although it is hard to access the neutron star regime
$\rho_n \ll \rho_s$ in He II, where interatomic forces are appreciable,
transitions between
turbulent and laminar superfluid vorticiy have been observed in
experiments with microspheres immersed in $^4$He at {\rm mK} temperatures
\citep{nks02}. 
Promising results on the relaxation of rotating He II were obtained by \citet{tsatsa80}, but
again these results are for $\rho_n \lsim \rho_s$ and hollow spheres
rather than a differentially rotating shell.
We propose to extend these experiments in two directions: (i) by
investigating low-Rossby-number ($\Delta \Omega/\Omega \ll 1$),
high-Reynolds-number ($\Rey \gg 10^5$) SCF with He II at the
temperature which minimizes $\rho_n/\rho_s$; and (ii) by repeating
(i) with a nonideal dilute-gas Bose-Einstein condensate confined
in a differentially rotating magneto-optical trap, in order
to probe the stability of a vortex lattice to Kelvin wave
excitations in the regime $\rho_n \ll \rho_s$ \citep{pa05}.
The presence
of a vortex tangle in He II can be detected by standard second-sound
absorption techniques \citep{hv56a,sbd83}, and
the torque in experiment (i) can be
monitored to look for 
oscillations when a change from HV to GM friction
(or vice versa) is triggered by the DGI. 

In classical Navier-Stokes fluids, injection of vorticity
into a metastable laminar state can trigger turbulence,
e.g. seed vortices injected  into a cylindrical vessel containing $^3$He-B 
(with $T \leq 0.6 T_c$) generate a vortex tangle
that  eventually decays into a rectilinear vortex array
\citep{finne03}. Unlike
He II, the normal component in $^3$He-B is laminar
in these experiments and does not participate in the turbulent dynamics,
due to its comparatively high viscosity
[$\nu_n \sim 1$ {\rm cm}$^2$ {\rm s}$^{-1}$ $\gg \nu_s$, cf. 
$\nu_n \sim \nu_s$ in He II; see \citet{fbek04}].
Standard glitch theories assume that the normal
component is tightly coupled to the crust
by the external magnetic field \citep{as88,jm98}. This suggest a second
possible SCF experiment with $^3$He-B (at $T \leq 0.6 T_c$) in a 
hollow spherical container in which the normal fluid
corotates with the container, and therefore does not participate
in the DGI, while the superfluid is free to become turbulent. 
Such an experiment can probe what aspects of the
turbulent-laminar transition are caused by the normal and
superfluid components respectively. Transitions
between a vortex tangle and a rectilinear array
can be detected using non-invasive nuclear magnetic
resonance techniques \citep{finne03}.

\acknowledgments We gratefully acknowledge
the computer time supplied by the Australian Partnership for
Advanced Computation (APAC) and the Victorian Partnership for
Advanced Computation (VPAC). 
We also thank S. Balachandar, from the University of
Illinois at Urbana-Champaign, for supplying us with his
original pseudo-spectral solver, designed for a single
Navier-Stokes fluid, from which our two-fluid, HVBK solver
was developed.
This research was supported by
a postgraduate scholarship from the University of Melbourne.


\newpage
\begin{figure*}
\caption{Streamlines in superfluid spherical Couette flow
with $\delta=0.4$, $\Rey=3 \times 10^4$, and $\Delta \Omega=0.1$.
(a) Normal fluid with GM mutual friction. (b) Superfluid
with GM mutual friction. (c) Normal fluid with HV mutual friction.
(d) Superfluid with HV mutual friction.
The streamlines are calculated by integrating the in-plane components
of the velocity fields in the plane $x=0$ at $t=18$.}
\label{fig:fig1}
\end{figure*}

\begin{figure*}
\epsscale{0.7}
\caption{Normalized counterflow velocity $(\vns)_z/v_{\rm DG}$ in
superfluid spherical Couette flow with $\delta=0.4$, $\Rey=3 \times 10^{4}$, and
$\Delta \Omega=0.1$ at $t=46$. (a) HV mutual friction. (b) GM mutual friction.}
\label{fig:fig2}
\end{figure*}

\begin{figure*}
\epsscale{1.0}
\caption{Meridional streamlines of the normal fluid (left) and
superfluid (right) components, for $\delta = 0.5$, after the outer sphere is accelerated instantaneously
from $\Omega_2 = 0.8$ at $t<20$ to $\Omega_2 = 1$ at $t\geq20$, and
the mutual friction is changed simultaneously from GM to HV.
The snapshots correspond to (a) $t=21$, (b) $t=22$, 
(c) $t=50$, (d) $t=100$, (e) $t=120$, and (f) $t=140$. Time is expressed 
in units of $\Omega_1^{-1}$.}
\label{fig:fig3}
\end{figure*}

\begin{figure*}

\caption{Meridional streamlines of the normal fluid (left) and
superfluid (right) components as a function of time for $\delta=0.5$, $\Delta \Omega = 0.2$, and GM mutual friction. There is no spin-up event at $t=20$, unlike in Figure \ref{fig:fig3}.
The snapshots correspond to (a) $t=18$, (b) $t=20$, (c) $t=50$, (d) $t=100$, (e) $t=120$, and (f) $t=140$.  Time is expressed in units of $\Omega_1^{-1}$.}
\label{fig:fig4}
\end{figure*}

\begin{figure*}
\caption{Fractional change in angular velocity of the
outer sphere, $\Delta {\Omega}/{\Omega} =
[{\Omega}_2(t) - {\Omega}_2(20)]/{\Omega}_2(20)$, as a
function of time, before and after a spin-up event at $t=20$ where
the mutual friction is changed instantaneously from GM to HV and $\Omega_2$
jumps according to $\Omega_2 = 0.9 \rightarrow 1$
(solid curve), $\Omega_2 =  0.8 \rightarrow 1$ (dashed curve),
and $\Omega_2 =  0.7 \rightarrow 1$ (dotted-dashed curve).
Time is measured in units of $\Omega_1^{-1}$. Dimensionless gap width: 
(a) $\delta=0.2$, (b) $\delta=0.3$, (c) $\delta=0.4$,
and (d) $\delta=0.5$.}
\label{fig:fig5}
\end{figure*}

\begin{figure*}
\caption{Fractional change in the angular acceleration of the
outer sphere, $\Delta \dot{\Omega}/\dot{\Omega} =
[\dot{\Omega}_2(t) - \dot{\Omega}_2(20)]/\dot{\Omega}_2(20)$, as a
function of time, before and after a spin-up event at $t=20$ where
the mutual friction is changed instantaneously from GM to HV and $\Omega_2$
jumps according to $\Omega_2 = 0.9 \rightarrow 1$
(solid curve), $\Omega_2 =  0.8 \rightarrow 1$ (dashed curve),
and $\Omega_2 =  0.7 \rightarrow 1$ (dotted-dashed curve).
Time is measured in units of $\Omega_1^{-1}$. Dimensionless gap width: 
(a) $\delta=0.2$, (b) $\delta=0.3$, (c) $\delta=0.4$,
and (d) $\delta=0.5$.}
\label{fig:fig6}
\end{figure*}

\begin{figure*}
\epsscale{0.8}
\caption{(a) Evolution of the $z$ component of the torque
on the outer sphere (multiplied by $10^4$) as a function of time for 
a superfluid with GM mutual friction (solid curve)
and a classical Navier-Stokes fluid (dashed curve), with $\delta=0.5$,
$\Rey = 3 \times 10^4$, and $\Delta \Omega =0.1$. The torque
is expressed in units of $\rho R_2^5 \Omega_1^2$ and
the time in units of $\Omega_1^{-1}$. (b)
Fractional change in the angular acceleration 
$\Delta \dot{\Omega}/\dot{\Omega} =
[\dot{\Omega}_2(t) - \dot{\Omega}_2(20)]/\dot{\Omega}_2(20)$
following the spin-up event $\Omega_2=0.9 \rightarrow 1.0$ at $t=20$ 
with $\delta=0.5$ and $\Rey = 3 \times 10^4$, in three cases:
GM $\rightarrow$ HV transition (dashed curve), superfluid with
GM mutual friction (solid curve), and
classical Navier-Stokes fluid (dashed-dotted curve).}
\label{fig:fig8}
\end{figure*}

\begin{figure*}
\epsscale{0.8}
\caption{Meridional streamlines for superfluid spherical Couette flow with
$\delta=0.5$ and $\Rey=3 \times 10^4$, obtained by integrating
the in-plane velocity components in the plane $x=0$.
(a) Viscous fluid at $t=20$,
with $\Omega_2=0.9$ and $\Omega_1=1.0$, and (b) at $t=22$,
after a sudden spin up of the outer sphere $\Omega_2=0.9 \rightarrow 1.0$.
(c) Viscous normal component at $t=20$,
with $\Omega_2=0.9$, $\Omega_1=1.0$, and GM friction, and (d)
at $t=22$,
after a sudden spin-up of the outer sphere $\Omega_2=0.9 \rightarrow 1.0$,
while simultaneously changing the friction from GM to HV.
(e) Inviscid superfluid component at $t=20$, with $\Omega_2=0.9$, $\Omega_1=1.0$,
and GM friction, and
(f) at $t=22$, after a sudden spin up of the outer 
sphere $\Omega_2=0.9 \rightarrow 1.0$, while simultaneously changing the friction 
force from GM to HV. Time is measured in units of $\Omega_1^{-1}$.}
\label{fig:figc}
\end{figure*}

\begin{figure*}
\caption{Turbulent-laminar vorticity transition during a glitch. Evolution 
of the torque on the outer sphere before and
after the outer sphere is impulsively accelerated from $\Omega_2=0.9$ to $\Omega_2=1.0$ 
at $t=20$, for $\delta=0.5$.
The angular velocity of the outer sphere is $\Omega_2=0.9$
at $ 0 \leq t < 20$, and $\Omega_2(t) = 1.0 + 0.001(t-20)$ during the
time interval  $ 20 \leq t \leq 120$.
The five meridional slices are contour plots
of $|(\vv_n)_z/v_{\rm DG}|$ at (a) $t=20$, (b) $t=22$, (c) $t=50$,
(d) $t=70$, and (e) $t=120 $. Dark regions
indicate where the DGI is active, i.e. $|(\vv_n)_z/v_{\rm DG}| \geq 1.0$,
white regions indicate where the DGI is not active, i.e. $|(\vv_n)_z/v_{\rm DG}| < 1.0$,
and the mutual friction is of HV form.}
\label{fig:fig9}
\end{figure*}

\end{document}